\documentclass[a4paper,twocolumn, accepted=2021-06-08]{quantumarticle}
\pdfoutput=1
\usepackage{amsmath}
\usepackage{amsthm}
\usepackage{amssymb}
\usepackage{hyperref}
\usepackage{color}
\usepackage{graphicx}
\usepackage{url}
\usepackage{color}
\usepackage{bbm}
\usepackage[numbers,sort&compress]{natbib}
\usepackage{comment}
\usepackage{graphicx}
\usepackage{bm}

\usepackage{array}
\usepackage{booktabs}
\newcolumntype{C}[1]{>{\centering\arraybackslash}m{#1}}
\AtBeginDocument{
\heavyrulewidth=.08em
\lightrulewidth=.05em
\cmidrulewidth=.03em
\belowrulesep=.65ex
\belowbottomsep=0pt
\aboverulesep=.4ex
\abovetopsep=0pt
\cmidrulesep=\doublerulesep
\cmidrulekern=.5em
\defaultaddspace=.5em
}
\newcolumntype{R}[1]{>{\raggedleft\arraybackslash}p{#1}}
\AtBeginDocument{
\heavyrulewidth=.08em
\lightrulewidth=.05em
\cmidrulewidth=.03em
\belowrulesep=.65ex
\belowbottomsep=0pt
\aboverulesep=.4ex
\abovetopsep=0pt
\cmidrulesep=\doublerulesep
\cmidrulekern=.5em
\defaultaddspace=.5em
}

\newcommand{\<}{\langle}
\newcommand{\e}{\varepsilon}

\renewcommand{\>}{\rangle}
\renewcommand{\(}{\left(}
\renewcommand{\)}{\right)}

\renewcommand{\v}[1]{\boldsymbol{#1}} 

\newcommand{\Z}{\mathbb{Z}}

\newcommand{\G}{\mathcal{G}}
\newcommand{\E}{\mathcal{E}}
\newcommand{\V}{\mathcal{V}}

\usepackage{xcolor}

\begin{document}
\title{QED driven QAOA for network-flow optimization}
\author{Yuxuan Zhang}
\affiliation{Center for Complex Quantum Systems, University of Texas at Austin, Austin, TX 78712, USA}
\author{Ruizhe Zhang}
\affiliation{Department of Computer Science, University of Texas at Austin, Austin, TX 78712, USA}
\author{Andrew C. Potter}
\affiliation{Center for Complex Quantum Systems, University of Texas at Austin, Austin, TX 78712, USA}

\date{}
\begin{abstract}
We present a general framework for modifying quantum approximate optimization algorithms (QAOA) to solve constrained network flow problems. By exploiting an analogy between flow-constraints and Gauss' law for electromagnetism, we design lattice quantum electrodynamics (QED)- inspired mixing Hamiltonians that preserve flow constraints throughout the QAOA process. 
This results in an exponential reduction in the size of the configuration space that needs to be explored, which we show through numerical simulations, yields higher quality approximate solutions compared to the original QAOA routine. We outline a specific implementation for edge-disjoint path (EDP) problems related to traffic congestion minimization, numerically analyze the effect of initial state choice, and explore trade-offs between circuit complexity and qubit resources via a particle-vortex duality mapping. Comparing the effect of initial states reveals that starting with an ergodic (unbiased) superposition of solutions yields better performance than beginning with the mixer ground-state, suggesting a departure from the ``short-cut to adiabaticity" mechanism often used to motivate QAOA.
\end{abstract}
\maketitle

\section{Introduction}\label{sec:level1}
Combinatorial optimization (CO) tasks present many classically-hard computational problems, and abound in practical applications from vehicle routing to resource allocation, job scheduling, portfolio optimization, and integrated circuit layout. Finding optimal solutions to many practically relevant classes of CO problems is an NP-complete task, which is effectively intractable for large problems. In the past decades, quantum computers promise tantalizing speedups on certain classically-hard computational problems, such as integer factoring~\cite{shor1994algorithms} and structured searching~\cite{grover1996fast}. Unfortunately, barring an upheaval of complexity theoretic dogma, quantum optimization algorithms are not expected to efficiently yield optimal solutions to NP-hard problems. However, for classical optimization on hard problems, one typically aims for reasonable but suboptimal approximations, and tremendous effort has been put into improving the quality of approximate solutions. In a similar vein, there is widespread hope that quantum-heuristics could yield better approximate solutions than their classical counterparts. 

This hope has been largely fueled by the introduction of the Quantum Approximate Optimization Algorithm (QAOA), a hybrid classical/quantum framework originally motivated as a variational spin-off of the Quantum Adiabatic Algorithm (QAA)~\cite{farhi2014}. QAOA consists of $p$-rounds of stroboscopic alternation between a classical cost Hamiltonian and a quantum mixing Hamiltonian, with time intervals for each evolution treated as variational parameters that are classically optimized. While it was initially suggested that even a single round ($p=1$) QAOA could provide a quantum-improvement over classical state-of-the-art~\cite{2014arXiv1411.4028F}, the quantum/classical gap was quickly closed~\cite{DBLP:journals/corr/BarakMORRSTVWW15}, and there is growing evidence~\cite{hastings2019classical,farhi2020quantum,farhi2020,bravyi2019obstacles} that $p$ must generically scale with the problem-size in order to achieve improved approximate solutions. Due to the difficulty of analyzing QAOA-performance at large-$p$, establishing rigorous evidence of quantum advantage remains elusive, and the practical value of QAOA will likely be decided empirically (like many successful classical heuristic methods). 

Making QAOA into a successful quantum heuristic will require advances in problem encoding, and algorithm efficiency. A key weakness of traditional QAOA is that many relevant CO problems impose constraints among variables, which are not respected by the QAOA heuristic. A typical approach to QAOA would be to map a CO problem into a binary integer linear program (BILP), whose objective function is mapped to an Ising-like spin model that can be implemented on quantum hardware. In this formulation, constraints are typically softly enforced by adding a term to the cost Hamiltonian that energetically penalizes constraint violations. This approach is frequently inefficient, as it can result in exploration of an exponentially-large (in problem size) set of infeasible (constraint-violating) configurations, which has been shown to dramatically hamper performance~\cite{wang2020x}. 

An alternative technique is to modify the QAOA procedure to automatically satisfy constraints throughout the algorithm. In \cite{hadfield2019quantum,wang2020x}, this approach was used to tackle graph-coloring problems (among others), where a number-conserving mixing Hamiltonian was designed to preserve a one-hot encoding structure. Due to the intimate connection between symmetries and conservation laws, this highlights a connection between physical symmetries and constraints in CO problems, and suggests that physics-inspired solutions may be fruitful. 

In this work, we exploit another common ``symmetry" found in physical systems: gauge-invariance~\footnote{Strictly speaking, gauge invariance is not an ordinary symmetry, however the analogy is frequently useful.}, to implement a constraint-satisfying mixer for network flow problems. Network flow problems are defined on graphs, where each link of a graph has a directed flow of ``goods'' that takes real or integer values. In practice, flow could represent an amount of vehicles, goods, communication packets, etc., being transported through the network. Real-valued flow problems tend to admit classically efficient solutions via linear programming, whereas multi-commodity integer flow problems are often classically hard. Integer flow problems have a wide array of applications from vehicle routing, traffic congestion minimization, and package delivery, to communication network optimization. Each of these problem formulations share a common constraint structure: the amount of flow entering a vertex must match the total outgoing flow, plus (minus) a fixed amount at certain source (sink) nodes. 

This flow structure is a discrete analog of Gauss law in electromagnetism: $\nabla\cdot \v{E}=\rho$, where $\rho$ is the charge density, if we re-interpret the electric field $\v{E}$ as a flow emanating out of a node, and the charge $\rho$ as the amount of sourced or sinked goods. The central idea of this paper will be to exploit this analogy to develop a lattice quantum electrodynamics (QED) inspired QAOA-mixer that automatically preserves network-problem flow constraints.

The structure of the paper is organized as follows: we briefly summarize QAOA from the generalized perspective advocated in~\cite{2014arXiv1411.4028F}, and review the structure of lattice-QED. We then establish a direct relationship to flow problems on finite-dimensional graphs, and define a constraint-preserving generalization of QAOA using a QED-style mixer. We numerically compare the performance of modified QED-QAOA and the original (X-mixer) QAOA on a (classically easy) flow maximization problem, and show that the quality of approximate solutions increases in a way that is consistent with exponential-in-problem size scaling. We then explore QED-mixer performance on classically-hard traffic congestion minimization problems, and study the behavior with increasing problem size and number of QAOA rounds. A key step in the algorithm is preparing an initial constraint-preserving state that is a quantum superposition including all constraint-preserving states. Unlike the original QAOA, where the X-mixer ground-state can be accomplished with a transversal set of single-qubit rotations, the QED-mixer ground-state is more complicated. We explore and compare multiple strategies for initial state preparation, and find, perhaps surprisingly, that the QED-mixer ground-state is not optimal, suggesting a departure from the adiabatic-algorithm reasoning often used to motivate QAOA.

\section{Quantizing network flow problems}
To set the stage, we briefly review the constraint structure of network flow problems, introduce the specific problem types that we will use to illustrate the QED-inspired QAOA approach, and describe an implementation of their cost function as a quantum Hamiltonian acting on qudits.

\subsection{Constraints in Flow Problems}
A flow problem is defined on a graph $\G$ with vertices $\V$ and edges $\E = \{(u,v)|~u,v\in\V\text{ are connected}\}$. Here $\G$ is required to be a directed graph by many versions of flow problems, such as max-flow problems, but can be undirected in other cases like EDP. We denote the total number of vertices as $|\V|$, and the number of edges as $|\E|$. On each edge, we define a flow: $f(u,v)\in \mathbb{F}$ taking value in some field $\mathbb{F}$, with $f(u,v)=-f(v,u)$ . To facilitate implementation on discrete-leveled quantum computing systems, in this paper we will specialize to integer flows of $k$-different commodities (i.e. $\mathbb{F}=\Z^{k}$). We define the vertex from which a commodity originates or terminates as a source or sink node respectively. We denote the sets of source and sink nodes as $\{s_i\}_{i=1}^{k}$ and $\{t_i\}_{i=1}^{k}$, and the amount of flow to be delivered for the $i^\text{th}$ source-sink pair as $d_i$. 

While there are a large variety of flow-problem formulations, they all share a common constraint structure. Namely, valid flows may begin and terminate only on source and sink nodes, respectively:

\begin{align}\label{eq:flow-cons}
\sum_{v:(u,v)\in\E} f_i(u,v) = d_i(\delta_{u,s_i}-\delta_{u,t_i})~~~\forall ~u\in \V.
\end{align}

\begin{figure}[h]
    \includegraphics[width=0.53\textwidth]{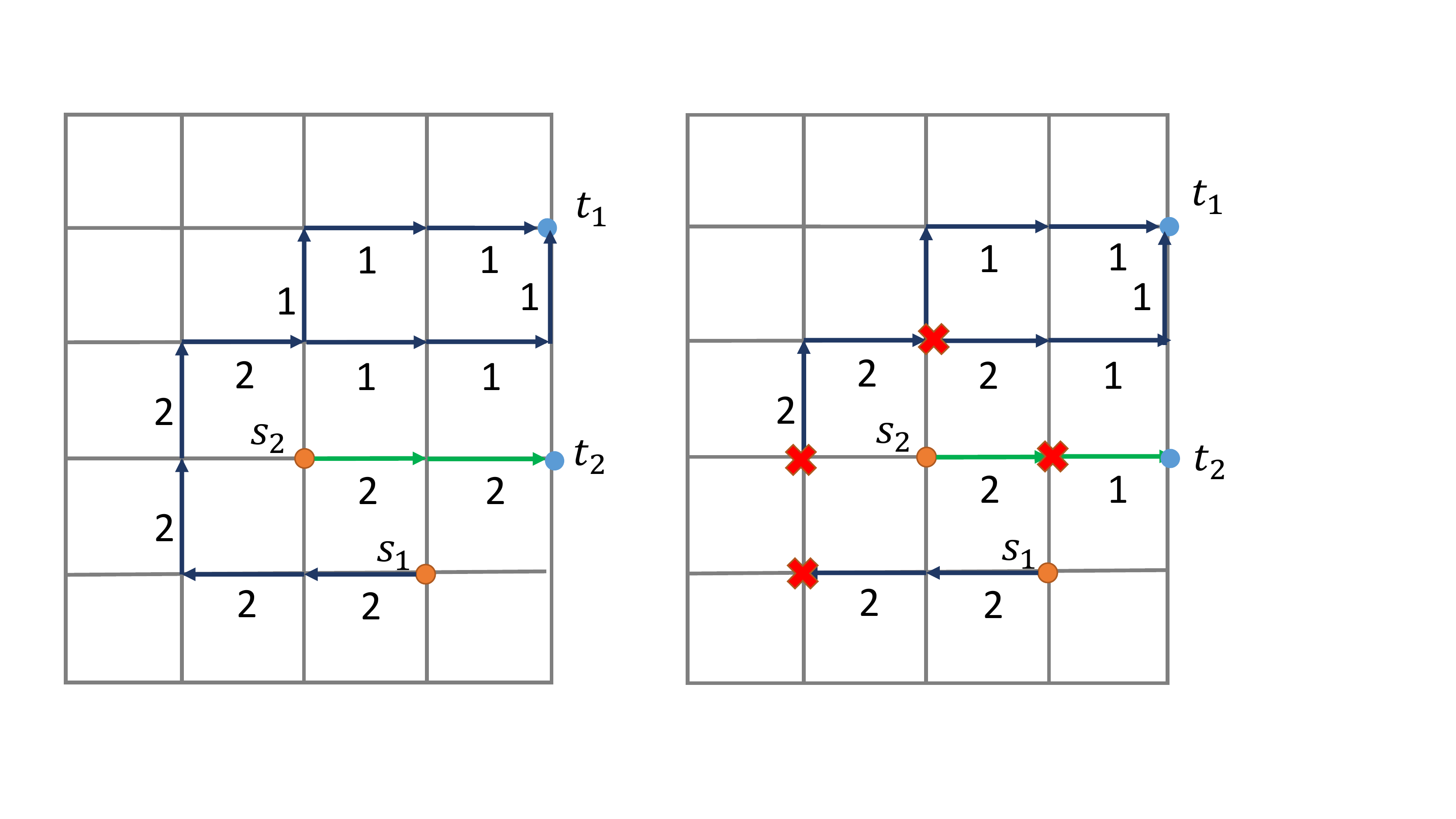}
    \caption{{\bf Example flows on a $5\times 5$ grid graph}   A feasible network flow configuration (L) and an unfeasible configuration(R): the arrows stand for flow directions, and different flows are distinguished by colors with numbers representing the amount of flow on each edge(a certain assignment of the flows in the graph is called a \emph{configuration}).}
    \label{fig:non-feas}
\end{figure}

Figure.~\ref{fig:non-feas} illustrates selected examples of valid and invalid flow configurations.

In addition, many flow problems impose additional capacity constraints on how many of each type of commodities may flow through a particular edge:
\begin{align}\label{eq:capacity-cons}
    \sum_{i\in [k]} |f_{i}(u,v)| \leq c(u,v)~~~\forall ~(u,v)\in \E,
\end{align}
where $c(u,v)\in \Z_+$ is referred to as the edge-capacity: the total amount of all types of flows cannot exceed the capacity on that edge.

Flow problems come in many varieties. Some, such as the single-commodity max flow problem, have efficient classical algorithms. However, many practical problems require introducing multiple commodities and imposing finite edge-capacities, which typically results in hard optimization problems. For example, the problem of maximizing capacitated integer flow was proven to be \textsf{NP}-complete even for only two source-sink pairs~\cite{eis75}.

\subsubsection{Qudit encoding}
To encode integer flow problems onto quantum hardware, we imagine using a register of $(2d_i+1)$-level qudits (possibly encoded into ordinary qubits using, e.g. binary or one-hot encoding) for each commodity and each edge $(u,v)\in\E$, with the qudit computational basis states $\{|-d_i\>,\dots,|-1\>,|0\>,|1\>\dots |d_i\>\}$
indicating the amount of flow on that link~\footnote{One could partially enforce capacity constraints by restricting the basis to $\{|-c(u,v)\>\dots |c(u,v)\>\}$ for each commodity, however, this choice would conflict with our scheme for producing a valid initial state.}
We note that, for this encoding the dimension of the entire Hilbert space is thus the same as the number of all possible configurations on the graph, which is $\prod_i (2d_i+1)^{|\E|}$. 

 The total Hilbert space of this encoded system contains exponentially many infeasible configurations that violate the flow constraints (Eq.~\eqref{eq:flow-cons}). The precise ratio of feasible (flow-conserving) to infeasible (flow-violating) solutions varies by graph; however, it is generally exponentially small in $|\V|$. To see this, note that, the distance between a pair of randomly chosen source and sink points is typically $\text{poly}(|\V|)$, and for each valid path from source to sink, removing any edge along the path from source to sink would result in an infeasible solution, resulting in combinatorially many infeasible solutions for each feasible one. 
 
\subsubsection{Flow operators}
We also introduce quantum flow operators on each edge $e\in \E$, and for each commodity type $i=1\dots k$:
\begin{align}
E^{(i)}_e = \sum_{f=-d_i}^{d_i} f|f\>\<f|_e \otimes \mathbbm{1}_{e'\neq e}
\end{align}
where the symbol $E$ anticipates an analogy with electric field operators in lattice-QED. Applying $E^{(i)}_e$ on a state would just return the amount of flow on edge $e$.

Furthermore, we denote the operator whose eigenstates are equal weighted superpositions of flow values as:
\begin{align}
X^{(i)}_e = \sum_{f,f'=-d_i}^{d_i}|f'\>\<f|_e \otimes \mathbbm{1}_{e'\neq e}
\end{align}
which is the natural qudit analog of the Pauli-X operator.

We also define the total flow of all goods on edge $e\in \E$, as $E_e\equiv \sum_{i=1}^k E^{(i)}_e$, and similarly $X_e \equiv \sum_{i=1}^k X^{(i)}_e$. The conventional QAOA mixer is built from $H_M = -\sum_e X_e$, which indiscriminately mixes between feasible and infeasible solutions, and has a tendency to get ``lost" in the exponentially larger infeasible parts of Hilbert space.

\subsection{The Edge-Disjoint Path Problem}
The main problem we will consider in this paper is a variant of traffic congestion minimization problem known as the edge-disjoint paths problem (EDP), often regarded as a particularly clean problem that characterizes the \textsf{NP}-hardness of flow optimization. 
Qualitatively, the frequently studied optimization version of EDP seeks to route $k$ different commodities without ``congestion", i.e., without multiple commodities flowing through the same edge: 
\\\\
{\bf EDP:} \emph{Given a undirected graph $G(V,E)$ with $k$ source/sink-pairs $(s_i,t_i)$, find $k$ paths connecting $s_i$ and $t_i$ for all $i\in [k]$ such that the maximum of congestion in each edge is minimized.}
\\

Since the maximum of congestion is a global function that is hard to implement on a quantum circuit, we can reformulate EDP's cost function by locally penalizing all congested edges instead:
\begin{align}\label{eq:relax_edp}
\min~~C \equiv \sum_{(u,v)\in E} \max\left\{0, ~\sum_{i\in [k]}|f_i(u,v)|-1 \right\}~~s.t.
 \nonumber\\
    ~\sum_{v:(u,v)\in\E} f_i(u,v) = d_i(\delta_{u,s_i}-\delta_{u,t_i})~~~\forall ~u\in \V.
\end{align}
In other words, in this version one aims at minimizing the total amount of congestion on all edges instead of the maximum. Notice that an optimal solution with $C=0$ will still be a solution of the EDP problem (with no congestion), whereas $C>0$ configurations may be regarded as approximate solutions of the relaxed EDP. 

EDP has been shown to be \textsf{NP}-hard even with a rather modest scaling of commodity types ($k\sim \log |\V|$)~\cite{cl12}. We restrict our attention to EDPs on planar graph, where the problem remains $\textsf{NP}$-hard ~\cite{cl12}.

To convert the EDP cost-function into a quantum Hamiltonian, we reformulate the cost function into an analytic form, and introduce the EDP cost Hamiltonian (using the encoding described above):
\begin{equation}\label{eq:costh_edp}
	H_\text{C,EDP} = \sum_{e\in E}[\frac{(2E_e^2-1)^2-1}{48}]
\end{equation}
which has vanishing energy for non-congested links (with $E_e=0,1$) and penalizes higher congestion. The normalization is chosen such that minimally congested links with $E_e=\pm 2$ have one unit of energy cost. Eq.~\eqref{eq:costh_edp} is a reformulation of the cost function in Eq.~\eqref{eq:relax_edp}, whose flow constraints will be dealt with in later sections.

\subsection{The Single Source Shortest Path Problem}
For classical simulations, the fully unconstrained multi-commodity Hilbert space quickly becomes intractable. Therefore, to benchmark the modified QAOA performance against the original formulation, we also consider a much simpler class of single source shortest path problem (SSSP), which seek the shortest path (on a weighted graph) between a single source and sink with unit demand ($d=1$):
\\\\
\noindent{\bf SSSP:} \emph{Given a weighted undirected graph $G(\V,\E)$, with weights $\{w_e: e\in \E\}$, and a single pair of source and sink vertices $s,t\in \V$, find the minimal length path connecting $s$ and $t$ where length is defined as the sum of the weights along the path.}
\\\\
Notice that SSSP can be defined on either directed, undirected or mixed graphs, and we choose to study the undirected version for consistency with the study of EDP. Efficient classical algorithms for SSSP \cite{bellman1958routing,dijkstra1959note} are textbook-standard materials (see also \cite{khadiev2019quantum} for a quantum algorithm for directed acyclic graphs). In this work, we do not aim to improve solution of SSSP, but only to use this problem as a benchmark to compare the performance of different QAOA mixers in the graph routing problem. Importantly, none of the QAOA strategies we test take advantage of the classically efficient solution, providing a fair comparison.

\begin{figure}[h]
    \centering
    \includegraphics[scale=0.35]{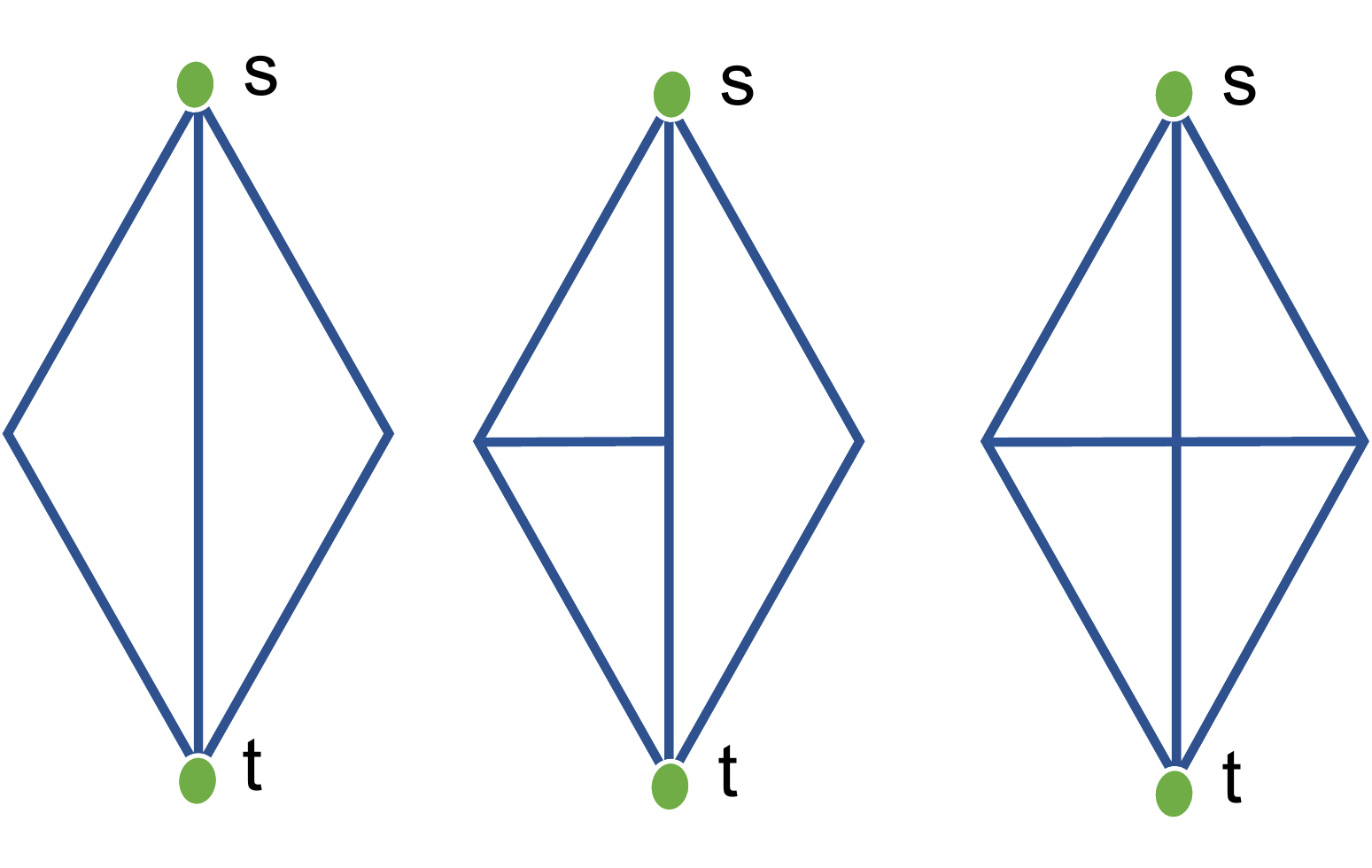}
    \caption{{\bf Triangle graphs used in the study of SSSP problem} In SSSP simulations we look for the shortest (lowest-weight) path from the top node to the bottom node, where weight on each edge is randomly assigned.}
    \label{fig:triangle}
\end{figure}

Since SSSP is a direct analogy of EDP (at $k=1$), we can use the same encoding scheme and write the cost Hamiltonian as
\begin{align}\label{eq:costh_sssp}
    H_\text{C,SSSP}=\sum_{e\in \E}w_{e}(E_e)^2
\end{align}
where $w_{e}$ denote the edge $e$'s weight.  Since only a single type of flow with demand 1 presents in the problem, the valid flow values are just -1,0,1 on any edge. The size of the Hilbert space of SSSP is thus $3^{|\E|}$, which, for large graphs, is far less than the $3^{k|\E|}$ (with $k\geq 2$) required for EDP, allowing us to classically simulate relatively larger instances.

With these problem classes in hand, we now turn to the task of modifying the QAOA algorithm to preserve the network flow constraints, beginning with a brief review of QAOA to set notation.
\renewcommand{\arraystretch}{1.2}
\begin{table}
\begin{tabular}{
>{\centering\arraybackslash}p{1.8cm} 
>{\centering\arraybackslash}p{1.8cm} 
>{\centering\arraybackslash}p{1.8cm} 
>{\centering\arraybackslash}p{1.8cm}}
 $\#$ of Triangles &  Total $\#$ States & $\#$ Feasible States  & Feasible fraction \\
 \toprule
 2 & 729 &3 &   $4.1\times 10^{-3}$\\
 3 & 2187 & 4 &$1.8\times 10^{-3}$\\
 4 & 6561 & 8&  $1.2\times 10^{-3}$\\
\end{tabular}
\caption{{\bf A comparison between total and feasible Hilbert space dimension} Total Hilbert space dimension ($|\mathcal{H}_\text{tot}|$) and feasible sub-space dimension ($|\mathcal{H}_\text{f}|$), and ratio of feasible to total states 
$|\mathcal{H}_\text{f}|/|\mathcal{H}_\text{tot}|$.
}
\label{tab:size}
\end{table}

\section{From QAOA to Lattice QED }
%
QAOA is designed to sample from low-cost states of a cost Hamiltonian $H_C$ which is diagonal in the computational basis, and represents the objective function of the optimization problem in question. In its original incarnation~\cite{2014arXiv1411.4028F}, the initial state $|\psi_0\>$, is chosen to be the ground-state of a mixing Hamiltonian $H_M = H_{M,X}$ with:
\begin{align} \label{eq:xmixer}
	H_{M,X} = -\sum_i X_i,
\end{align}
which we will refer to as the ``X-mixer." Subsequent generalizations~\cite{hadfield2019quantum} considered more complicated forms of $H_M$ designed to preserve constraints of various forms. The algorithm proceeds by evolving:
\begin{equation}\label{eq:qaoa}
	|\psi_p(\v{\gamma},\v{\beta})\>=\prod_{j=1}^p e^{-i\beta_{j}H_M}e^{-i\gamma_jH_C}|\psi_0\>
\end{equation}
to generate a variational wave function characterized by real-parameters $\{\gamma_i\}$ and $\{\beta_i\}$ $(i = 1, 2, ...p)$, which are classically optimized (using the classical routine of ones choice) to minimize the expected cost: $\v{\gamma}_*,\v{\beta}_* = \text{arg}\min \e_C$, where:
\begin{align}\label{eq:expected_cost}
	\e_C\equiv \<\psi_p(\v{\gamma},\v{\beta})|H_C|\psi_p(\v{\gamma},\v{\beta})\>.
\end{align}
This biases the wave-function amplitude of $|\psi_p(\v{\gamma}_*,\v{\beta}_*)\>$ towards low-cost configurations, such that repeated sampling from this state preferentially yields low-cost solutions.

In the limit of infinite $p$, QAOA contains Quantum Adiabatic Algorithm (QAA) as a subset of possible solutions and is guaranteed to find the exact optimum. For hard problem instances, precisely following the adiabatic path may require $p$ to grow super-exponentially with problem-size, but it is hoped that approximate short-cuts to this adiabatic solution may be variationally identified with far lower $p$. 

To apply this formalism, one must first map the optimization problem variables onto qubits, such that the cost for each qubit configuration can be computed in a local manner. For constrained optimization problems, this often results in a wasteful encoding in which many qubit states do not satisfy the feasibility constraints. One possible strategy would be to energetically penalize the constraint violation by introducing a penalty term into $H_C$ for unsatisfied constraints. While straightforward in its implementation, this strategy results in wasteful exploration of (typically exponentially many) configurations corresponding to infeasible solutions, degrading algorithm performance. An alternative option~\cite{hadfield2019quantum} is to identify an alternate mixing term $H_M$ which automatically preserves the constraint structure. Then, if an initial state can be prepared to satisfy all constraints, the algorithm will only search inside the feasible subspace. In what follows, we focus on the constraints common to a large variety of network flow problems and show how to encode them into an appropriate constraint-preserving mixer inspired by lattice-QED.

\subsection{Lattice QED Hamiltonian}
The flow constraints described in Eq.~\eqref{eq:relax_edp} are of precisely the same form as Gauss's law for lattice-QED, if we interpret each commodity flow as a different ``flavor" of electric field, and the corresponding sources and sinks as positive and negative charge $d_i$. This suggests that we can use gauge-invariant lattice-QED Hamiltonians to implement constraint-preserving mixers for the network flow QAOA. Here, we briefly review some relevant lattice-QED notations and formalisms. In what follows, we specialize them to planar graphs, although our construction generalizes to arbitrary finite-dimensional graphs (but would become infeasible for fully-connected graphs). For notational simplicity, we initially suppress the commodity (``flavor") label.

The Hamiltonian formulation of the (compact) lattice QED, on a planar graph $\G = (\V,\E)$, is defined by introducing discrete analogs of the continuum electric field $\v{E}(\v{r})$ and vector potential $\v{A}(\v{r})$. Specifically, a (gauge-redundant) Hilbert space is defined by electric field operators $E_{uv}=-E_{vu}$ on each edge $(u,v)\in \E$ whose eigenstates are denoted $|e_{uv}\>$ with $e_{uv}\in \Z$. Electric fields are oriented such that $E_{vu}=-E_{vu}$. The conjugate operator to $E_{uv}$ is denoted by $e^{-iA_{uv}}$, which raises or lowers the electric field: 
\begin{align}\label{eq:efield-raising}
	e^{iA_{uv}}E_{uv}e^{-iA_{uv}} = E_{uv}+1,
\end{align}
and $[e^{-iA_{uv}},E_{wx}]=0$ for $(w,x)\neq(u,v)$.

Physical states are defined by projecting onto subspace that satisfies a lattice analog of the continuum Gauss' law $\nabla\cdot \v{E}(\v{r})=\rho(\v{r})$, i.e. $\sum_{u:(u,v)\in \E} E_{uv}=\rho_u$, which is precisely the same form as the flow constraint (Eq.~\eqref{eq:flow-cons}), provided that we equate the electrical charge with demand, $d$. The Gauss' law is equivalent to demanding invariance under gauge transformations:
\begin{align}
e^{-iA_{uv}}\rightarrow e^{-i\phi_u}e^{-iA_{uv}}e^{i\phi_v}\\
|\psi\> \rightarrow e^{i\sum_{u\in \V} \phi_u \rho_u} |\psi\>
\end{align}
for any vertex-dependent phases $e^{i\phi_v} \in U(1)$.

A special role is played by gauge invariant, Wilson loop operators, $U_\Gamma = e^{-i\oint_\Gamma \vec{A}\cdot d\vec{\ell}}$, which measure the magnetic flux through a closed oriented loop $\Gamma$, where we use integral notation to indicate the product of $e^{-iA_{uv}}$ over all links $(u,v)$  on the perimeter of $\Gamma$, with orientation along that of $\Gamma$. On planar graphs, which have trivial homology, an arbitrary Wilson loop can be decomposed into a product of small loop operators circling the elementary faces (plaquettes) of the graph, which we label by $\mathcal{F}$. 

For dimensions $d>2$, ordinary Maxwell electrodynamics emerges as the continuum and weak-coupling limit of the minimal gauge invariant Hamiltonian:
\begin{align}
	H_\text{Maxwell} = \frac{K}{2}\sum_{uv\in \E} E_{uv}^2-\sum_{f\in \mathcal{F}}(U_f+U^\dagger_f)
\end{align}
where $U_f$ denotes the Wilson loop encircling face $f$ in the right-handed sense, and $K$ is a coupling constant. The first term represents an electric field line tension, whereas the second gives an energy cost to magnetic flux (which produces quantum dynamics for electric fields). For $d=2$, the lattice-QED systems is confined by monopole/instanton proliferation for any non-zero electric field line tension, $K>0$.

\subsection{QED-Mixer for Network flow problems}
To obtain a flow-conserving mixer, one can nominally choose any gauge-invariant lattice-QED Hamiltonian, replacing electric field variables with flow variables. We introduce a separate electric-field ``flavor" for each type of commodity indicated by a superscript parenthetical index: $E^{(i)}$ with $i=1\dots k$. In practice, we will choose our mixing Hamiltonian as the minimal Maxwell Hamiltonian, since it contains only the minimal elementary Wilson loops, thereby simplifying its implementation. Furthermore, we will set the electric field tension $K$ to zero, since the goal of a mixer Hamiltonian is to produce unbiased quantum tunneling between different flow configurations. Significant efforts have been devoted to developing various schemes for ``qubitization" and quantum simulation of lattice gauge theories~\cite{Marcos_2014}. We will remain largely agnostic about the specific implementation details, however, it is crucial to truncate the range of electric field values to lie between $-c(u,v)\leq E_{u,v}\leq c(u,v)$. To this end, we modify the electric field raising operator $e^{-iA_{uv}}$ to annihilate $|c(u,v)\>$, without altering its action on other states. We refer to the resulting Hamiltonian:
\begin{align}\label{eq:naive_mix}
	H_\text{M,QED} = -\sum_{i=1}^{k}\sum_{f\in \mathcal{F}}(U^{(i)}_f+\text{h.c.})
\end{align}
as the QED-mixer. We require that sufficiently many elementary faces/plaquettes $f\in\mathcal{F}$ are included to provide a complete basis of graph cycles, so that evolution under $H_m$ can transfer any flow-configuration to any other flow configuration. This is easy to satisfy for planar  graphs, one can readily verify that $O(|\V|)$ applications of $H_m$ connect any any two flow configurations (see Appendix~\ref{app:completeness}). We note that the circuit complexity of implementing this mixing Hamiltonian grows length with number of minimal cycles.

\subsubsection{Avoiding Isolated Loop Generation}
\begin{figure}[tb]
    \centering
    \includegraphics[width=0.25\textwidth]{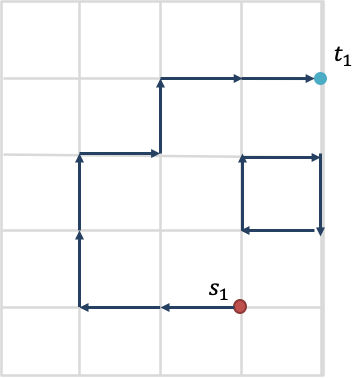}
    \caption{{\bf A configuration with an isolated loop} Without the loop which is detached from the path, this would be a feasible solution. One could remove it easily, but having multiple isolated loops in a complicated graph would make such process hard to perform}
    \label{fig:isoloop}
\end{figure}

\begin{figure}[tb]
    \centering.
    \includegraphics[width=0.42\textwidth]{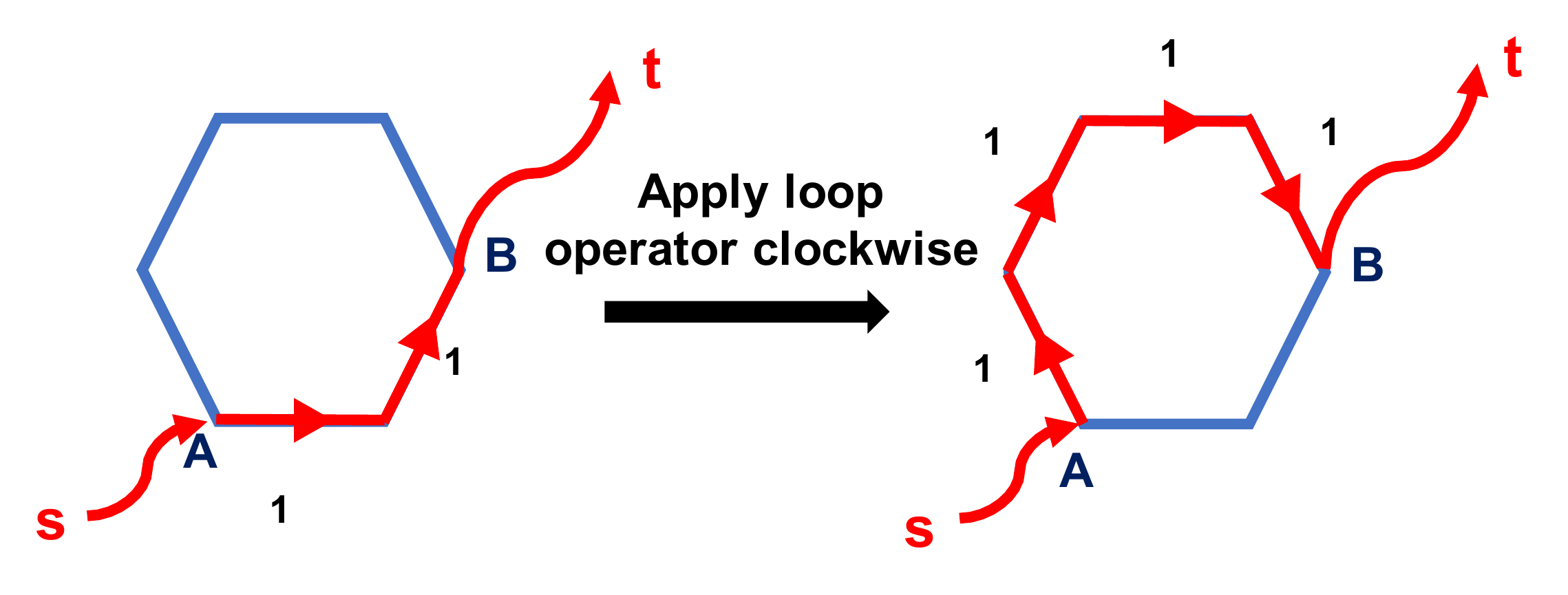}
    \includegraphics[width=0.5\textwidth]{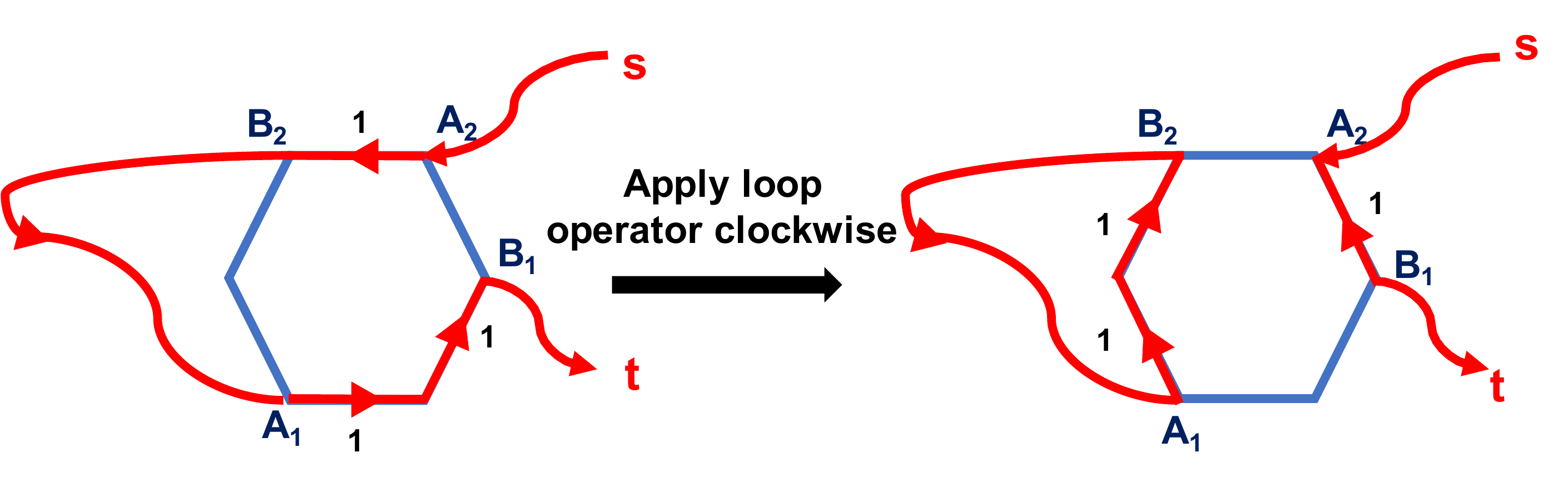}
    \caption{{\bf An explanation of the ``decision function"} For simplicity, we consider only one type of flow with max capacity 1. In both pictures, a flow (marked red) initially travels through the plaquette and then a loop operator is applied, increasing the flow on each edge on the plaquette by 1 in clockwise direction. The only difference between the pictures is that, the flow enters the loop twice at $A_1$ and $A_2$, and applying the operator resulted in redirected flow from $A_1$ to $B_2$, resulting in an isolated loop $A_1, B_2, ..., A_1$. To avoid such instances, we only apply the loop operator when exactly one continuous path of flow appears in the plaquette, which can be determined locally.}
    \label{fig:iso}
\end{figure}

As written, the QED-mixer does not allow any flow constraint violations. However, this mixer still suffers from a potential problem: it can crate isolated loops of circulating flow that do not connect to sources or sinks (see Figure.~\ref{fig:isoloop}). These isolated loops satisfy all flow constraints, but do not correspond to a physically relevant solution.
One option is to simply retain these isolated loops throughout the QAOA, and prune them from the final solutions via classical post-processing. A potential drawback is that is that isolated loops may incur unphysical costs, and on large graphs, each valid path can be dressed with exponentially many isolated loops, each of which could incur an unphysical cost penalty, masking the true cost of the ``pruned" post-processed solution during the QAOA optimization. 
Throughout the remainder of this paper, we will restrict our attention to problems with unit demand for each type of good. For this subclass of problems, we can avoid isolated loop creation by introducing further restrictions on the QED-mixer, which we call the restricted QED (RQED) mixer. In practice, this restriction will incur additional circuit complexity and may be undesirable. We will later compare the performance of the QED-mixer with and without restriction.
The key step will be formulating a method to efficiently detect whether acting with $U_f$ or $U_f^\dagger$ would create an isolated loop, depending on the graph property and specific problem. To avoid combinatorial blow-up of Hamiltonian terms, this detection must be done locally, which we do as follows. To determine whether adding electric field circulation around an elementary cycle of the graph adds an isolated loop, consider acting with $U_f^\dagger$ to add an electric field loop to a simple path and the following steps: Traverse the edge segments of the cycle in a counterclockwise fashion. For each vertex $v\in \V$, count the number of electric field lines entering ($E^{(i)}_{v,\text{in}}$) versus leaving ($E^{(i)}_{v,\text{out}}$) the node. Denote their difference-squared as 
\begin{align}
\mathsf{V}^{(i)} \equiv \sum_{j=1}^\ell \left(E^{(i)}_{v_j,\text{in}} - E^{(i)}_{v_j,\text{out}}\right)^2,
\end{align}
where $v_1,\dots, v_\ell$ are the nodes in the cycle. Notice that $(E^{(i)}_{v,\text{in}} - E^{(i)}_{v,\text{out}})^2$ can only take value 1 or 0. Since in our setting where maximum flow is 1, having two different direction of flows at the same node would suggest the node being used repeatedly, which further implies the configuration already contains an isolated loop. Imposing the Gauss' law constraint, $\mathsf{V}^{(i)}$ is equal to the total number of electric field lines entering or exiting the loop (if the loop does not contain a source/sink; or one could interpret a source as outside flow entering the loop and vice versa) without regard to sign (which is necessarily even). One can readily check that an isolated loop will be created unless $\mathsf{V}^{(i)}=2$ (see Figure.~\ref{fig:iso} for sample instances).

With this in mind, we can then left-multiply $U_f^\dagger$ by a locally evaluable ``decision function" to define a modified loop operator:
\begin{align}\label{eq:restricted_QED}
	U_f^{(i)}\rightarrow \tilde{U}_f^{(i)}=\delta_{\mathsf{V}^{(i)},2}U_f^{(i)},
\end{align}
which does not create isolated loops. Note that $\delta_{\mathsf{V}^{(i)},2}$ commutes with $U$ so the multiplication order is arbitrary.

In practice, $\delta_{\mathsf{V}^{(i)},2}$ can be written as a polynomial with zeros at all even values of $\mathsf{V}^{(i)}$ other than $2$: 
\begin{align}
	\delta_{\mathsf{V}^{(i)},2}= \prod_{j=0,1\dots \ell; j\neq 1} \left(\frac{2j-\mathsf{V}^{(i)}}{2j-2}\right),
\end{align}
which permits implementation with circuit complexity $\sim\text{poly}(\ell)$. For simple graph structures, such as grids, where the sizes of elementary cycles are bounded independent of the system size, imposing this restriction adds only constant circuit-depth overhead.

 

\subsubsection{Initial State Preparation}\label{subsec: state_prep}
To begin the QAOA procedure, one must choose an initial state that is a quantum superposition with weights on all possible solutions.
In the original formulation of QAOA, the initial state was chosen as the ground-state of the X-mixer Hamiltonian. This had two virtues: first, it ensured that QAOA could reduce to the quantum adiabatic algorithm in the limit of large step number, $p$. Second, this state is an equal weighted superposition of all computational states, and does not introduce an intrinsic bias.

In contrast, for QED-mixers, the mixer ground-state is no longer an equal-weight superposition. Moreover, it is not straightforward to implement the ground-state of the QED or RQED mixers. For these reasons, we consider alternative state preparation schemes. As a starting point, we assume that it is straightforward to greedily prepare a computational basis state that satisfies the flow-constraints (a detailed prescription will be given below for EDP problems). 

\paragraph{Adiabatic ground-state preparation by reverse-annealing:} One option would be to adiabatically prepare the QED or RQED mixer ground-state via adiabatic evolution from a classical Hamiltonian with the fixed computational basis state as the ground-state to the (R)QED mixer ground-state. However, generically, the QED mixer will have gapless photon-like excitations, whose gap scales to zero as $\sim 1/R$ where $R$ is the graph radius (maximal distance between two nodes), such that this adiabatic ground-state preparation requires time $\sim \mathcal{O}(R)$. Moreover, we will see that starting from the ground-state of the mixer Hamiltonian actually leads to worse QAOA performance, due to the reasons we will discuss in later sections.

\paragraph{State preparation by mixer evolution:} An alternative approach is to simply time-evolve the initial flow-constraint-preserving computational basis state with the mixing Hamiltonian for a certain amount of time, which spreads out the weight  of the Hamiltonian onto other configurations. In analogy to photon propagation in electrodynamics, the flow should spread out ballistically (moving with constant velocity), covering the graph in time $\sim O(R)$. Hamiltonian simulation techniques can implement time-evolution for time $t$ with performance that asymptotically tends to $O(t)$~\cite{Berry_2015}. In practice, it may not be necessary to simulate continuous time evolution, but rather one could break $H_M$ into local terms acting on disjoint sets of qubits and stroboscopically alternate among them to achieve similar results. 

To numerically analyze the spreading of the wave function, we introduce the inverse participation ratio (IPR) test:
\begin{equation}
	\text{IPR} = \sum_i|\psi_i|^4,
\end{equation}
where $\psi_{i}$ is the amplitude of the wave-function in computational basis state $i$. IPR measure is inversely proportional to how evenly the wave-function spread-out over the computational basis states (i.e., among potential solutions to the optimization problem). The choice of power 4 here is because 2 would always give 1 and higher powers contain the same information about the wave function as 4th power does except for rare cases. When the wave-function is concentrated on a single state, $\text{IPR}=1$; whereas an equal superposition of all states yields the minimal value of $\text{IPR}=1/|\mathcal{H}|$, where $\mathcal{H}$ is the size of the Hilbert space (number of feasible solutions)
\begin{figure}[h]
    \centering
    a) \includegraphics[width=0.48\textwidth]{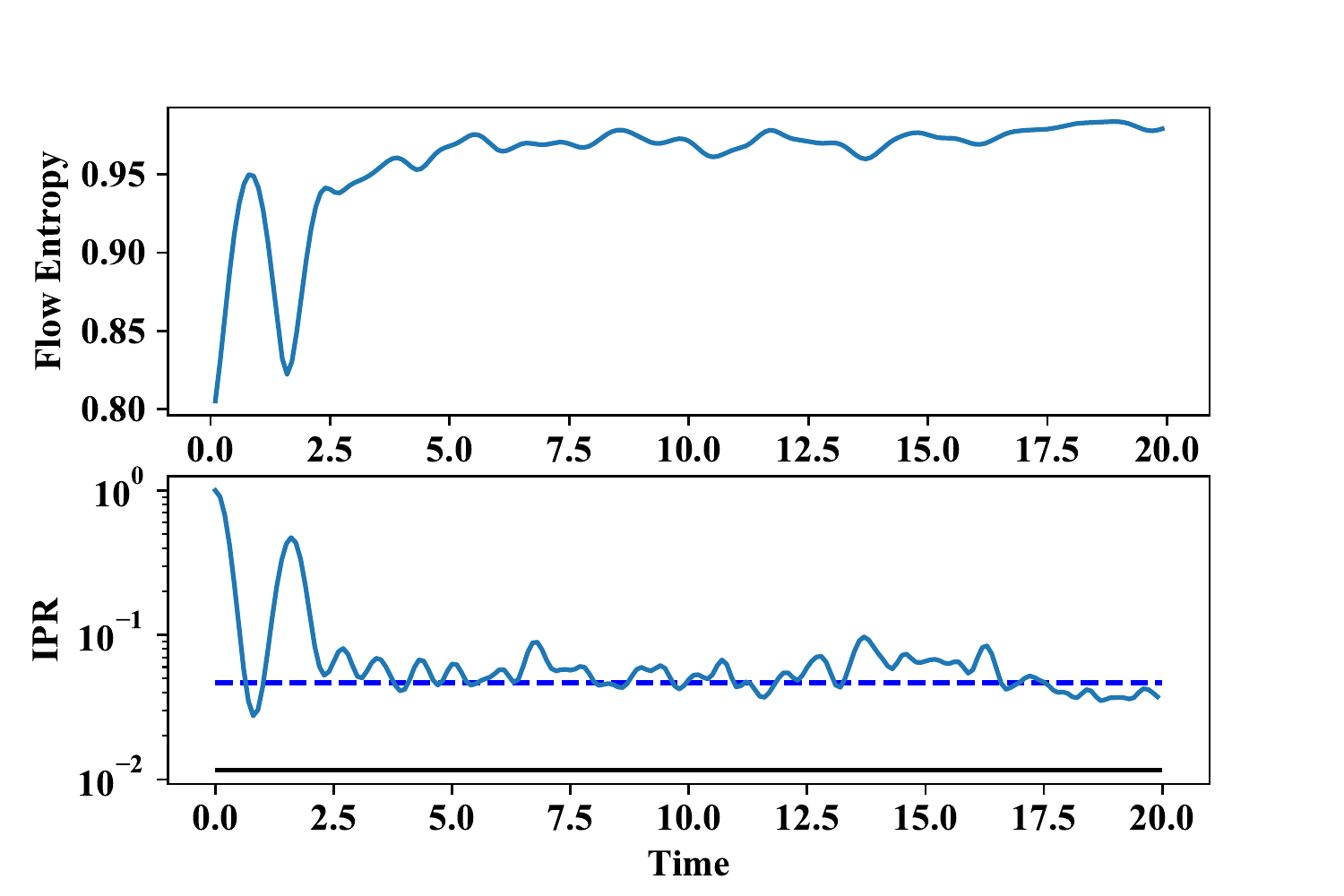}
    b) \includegraphics[width=0.48\textwidth]{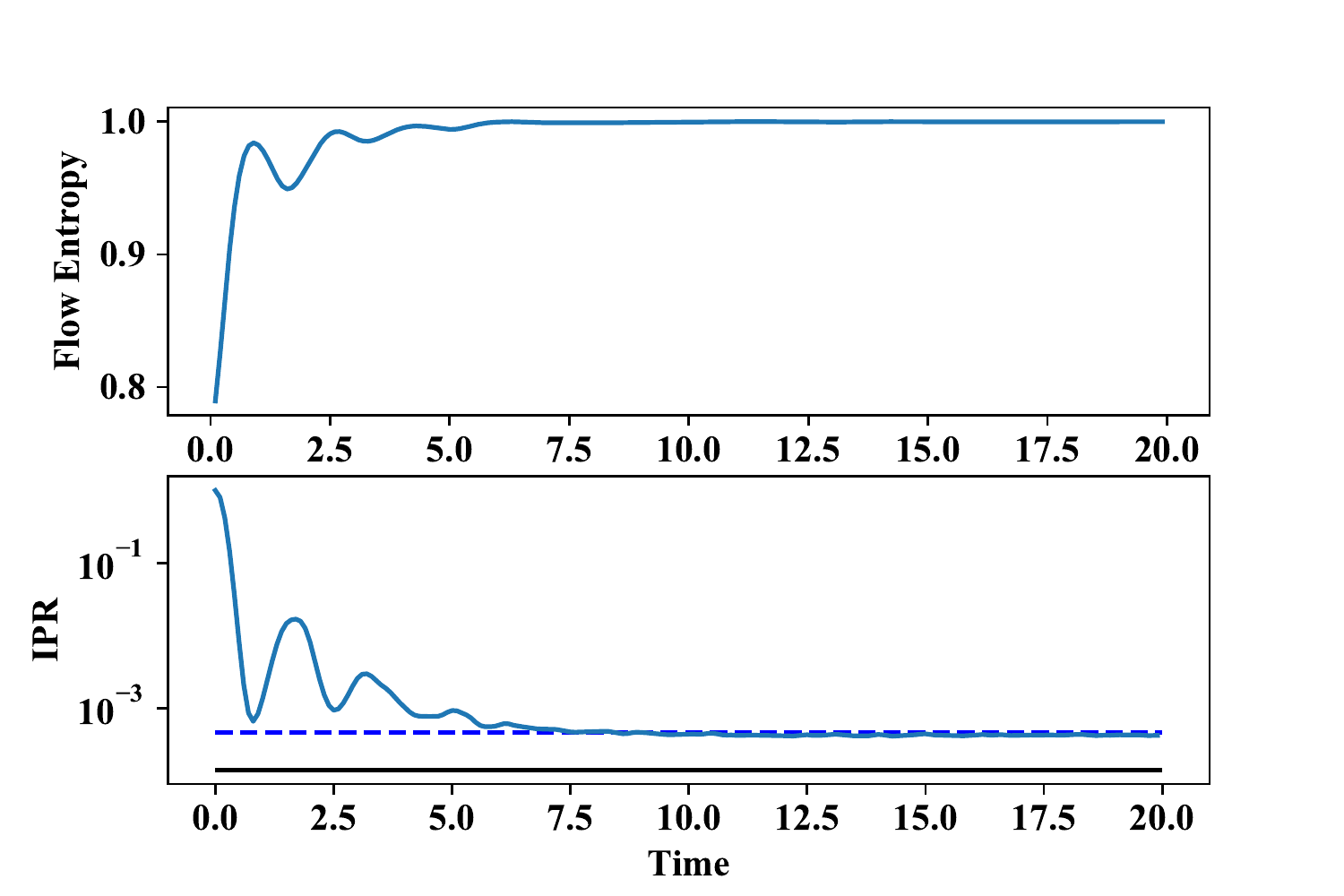} 
    \caption{{\bf IPR and entropy test -- } The figures show instances of IPR and flow entropy (normalized to its maximum) $S$ in the feasible space for single-source $4\times4$ (top) and $5\times5$ (bottom) square lattices for RQED-mixer in real time evolution. In IPR tests, the black solid line stands for the minimum possible IPR value (equal superposition of all possible paths from $s$ to $t$), and the blue dashed line shows the IPR for mixer ground state. All $s-t$ pairs and initial paths are drawn at random. At both sizes, the entropy curve characters the bumps and saturation in the IPR curve, suggesting itself as a good alternative of IPR.
    \label{fig:IPR}
    }
\end{figure}

Figure.~\ref{fig:IPR} shows the evolution of IPR with evolution under the RQED-mixer for single source-sink pairs on different sized square-grids. Since the RQED-Hamiltonian only evolves in the feasible solution space the test is done only within a constructed feasible subspace. The IPR decays from one, approximately saturating to a value close-to, but below the IPR for the mixer ground-state (blue dashed line), in characteristic time $t_\text{sat}\sim O(R)$ (where $R$ is the graph radius). In addition, the IPR exhibits approximately periodic revival behaviors, which are most evident on the smaller $4\times 4$ grid. As is well known from the study of Poincare recurrences, the period of these revivals becomes (doubly)-exponential in size of the graph, since the number of feasible solutions grows exponentially with the size of the graph, and can be safely neglected even for moderate graph sizes (indeed the oscillations are negligible already for the $5\times 5$ grid.)
 
  To prepare the initial state for subsequent QAOA iterations, we evolve the state until it just enters the saturation region where the IPR stabilizes to its long time value (e.g. in the $5\times5$-grid this occurs around $t_\text{sat}\approx 7.5$, see Figure.~\ref{fig:IPR}). In both tests, we observe that, inside the saturation region, the saturation-value of IPR lies below that of the mixer's ground-state, indicating that the mixer ground-state is more biased than the time-evolved state. This feature is natural since the evolved state is not low-energy and can be expected to contain additional configurational entropy.

In practice, IPR is challenging to measure as the Hilbert space size grows exponentially. Instead, one can determine the saturation time by monitoring local observables that act as witnesses for the IPR. Without loss of generality we focus on a single commodity case, since for multiple commodities, the Hilbert space is a tensor product of the single-commodity Hilbert spaces, with no inter-commodity interactions in the state preparation procedure. We examine the probability of observing unit flow (of either sign) on edge $e\in \E$ after evolution for time $t$ under the mixing Hamiltonian
\begin{align}\label{eq:probability}
    p_e(t) =\frac{\<E_e^2\>}{\sum_{e\in \E}\<E_e^2\>}
\end{align}
which can be estimated by sampling from the state in the computational basis.
 
We then define the (normalized) ``flow entropy" as the von-Neumann entropy of this probability distribution:
\begin{align}\label{eq:entropy}
    S_f= -\frac{1}{|\mathcal{E}|\log 2} \sum_{e\in \E}p_e\log(p_e).
\end{align}
Larger $S_f\leq 1$ represents a more even distribution of paths. $S_f$ saturates its maximal value of $1$ when each link carries flow with equal probability. The flow entropy exhibits similar saturation behavior to the IPR, allowing one to measure the saturation time for a given graph. Crucially, to accurately estimate flow, the probability $p_e$ needs only be measured to accuracy $\sim 1/|\E|$, which requires sampling cost $\sim |\E|^2$ that is polynomial in problem size (in contrast to the exponentially small IPR), allowing an efficient measurement to identify saturation time at which to stop the state preparation step.

\subsection{Algorithm description}
We are now ready to detail the steps of the modified QAOA for network flow problems. Given a directed graph $G(\V,\E)$ as input (if the graph is undirected, simply choose an arbitrary orientation for the edges):
\begin{enumerate}
\item \emph{Pre-process:} 
Identify a set of elementary faces (i.e. choose a basis of closed cycles) in $G$ and store them. For a planar graph, this can be done classically in polynomial time \cite{BCKO08}. 

\item \emph{Hamiltonians simulation:}
Choose a technique to simulate time-evolution under  the cost and mixing Hamiltonians: $H_C$, $H_M$.

\item \emph{Initial state preparation:} 
As described in Section~\ref{subsec: state_prep}, for each pair $(s_i, t_i)$ given in the input, pick an arbitrary ``seed" path, $P_0$ (which can be found efficiently by standard methods), and define the corresponding computational basis state as $|P_0\>$. Identify the saturation time $t_\text{sat}$, for the graph by the flow-entropy test described in the text. Then, simulate time-evolution under the mixing Hamiltonian to form the initial state: $|\psi_0\> = e^{-iH_Mt_\text{sat}}|P_0\>$.

\item \emph{Variational Optimization:}
Following the original QAOA procedure, but  replacing the the X-mixer with the (R)QED-mixer to avoid generated flow-constraint violations, find $\v{\gamma}_*,\v{\beta}_* = {\rm arg} \min \e_C$ using any desired classical minimization procedure, 

\item
\emph{Post-process} 
Repeatedly sample from the optimized variational state $|\psi(\v{\gamma}_*,\v{\beta}_*)\>$, recording the best (lowest-cost) sample encountered as an approximate solution.
\end{enumerate}

\section{Numerical simulation of algorithm performance}
In this section we present results from numerical simulation of QED-modified and standard QAOA of small-scale network flow problems. Due to the rapid growth of Hilbert space, $|\mathcal{H}|\sim O(3^{k |\E|})$, the accessible problem size is quite limited. In order to provide a meaningful comparison of the QED-mixer, we first consider the (classically easy) SSSP problem ($k=1$), which will allow simulation of relatively larger graphs to enable a comparison of QED-mixer and X-mixer. We then simulate EDP problems with $k=2$ on a grid graph for RQED-mixer only, where we can restrict our numerical simulation to the feasible solution space of size $\ll |\mathcal{H}|$. 

For the original X-mixer, in each step, the variational parameters can be limited to $[0, 2\pi]$ for $\{\gamma_i\}$ and $[0, \pi]$ for $\{\beta_i\}$, due to the periodicity of evolution under Pauli strings. The QED-mixer has no such periodicity. However, to run the QED-mixer for longer times would require additional circuit depth with which additional rounds of QAOA with the X-mixer could have been performed. Hence, to make a fair comparison, we also restrict our variational parameter ranges for the QED-mixer to the same range as for the original X-mixer.

In all simulations, we first perform a global search with differential evolution, and then optimize with a local BFGS method~\cite{fletcher2013practical}. For both methods, we restrict the optimizer to at most 200 minimization steps to balance accuracy and efficiency. 

To generate a larger collection of problems from a limited set of graph types and sizes, we generate random problem instances for each graph. For the SSSP problems we consider for each triangle graph in Figure.~\ref{fig:triangle} with source-and-sink located at opposite corners, we generate random problem instances by drawing random weights $w_e$ i.i.d. for each edge from the uniform distribution on the unit interval $[0,1]$, and seeding the state-preparation step with a uniformly-randomly chosen path, $|P_0\>$. For the EDP problem, the edges are unweighted, so we further choose the source and sink locations uniformly at random on different sized grid graphs.

\subsection{Comparing Mixers}
\begin{figure}[t]
    \centering
    \includegraphics[width=0.5\textwidth]{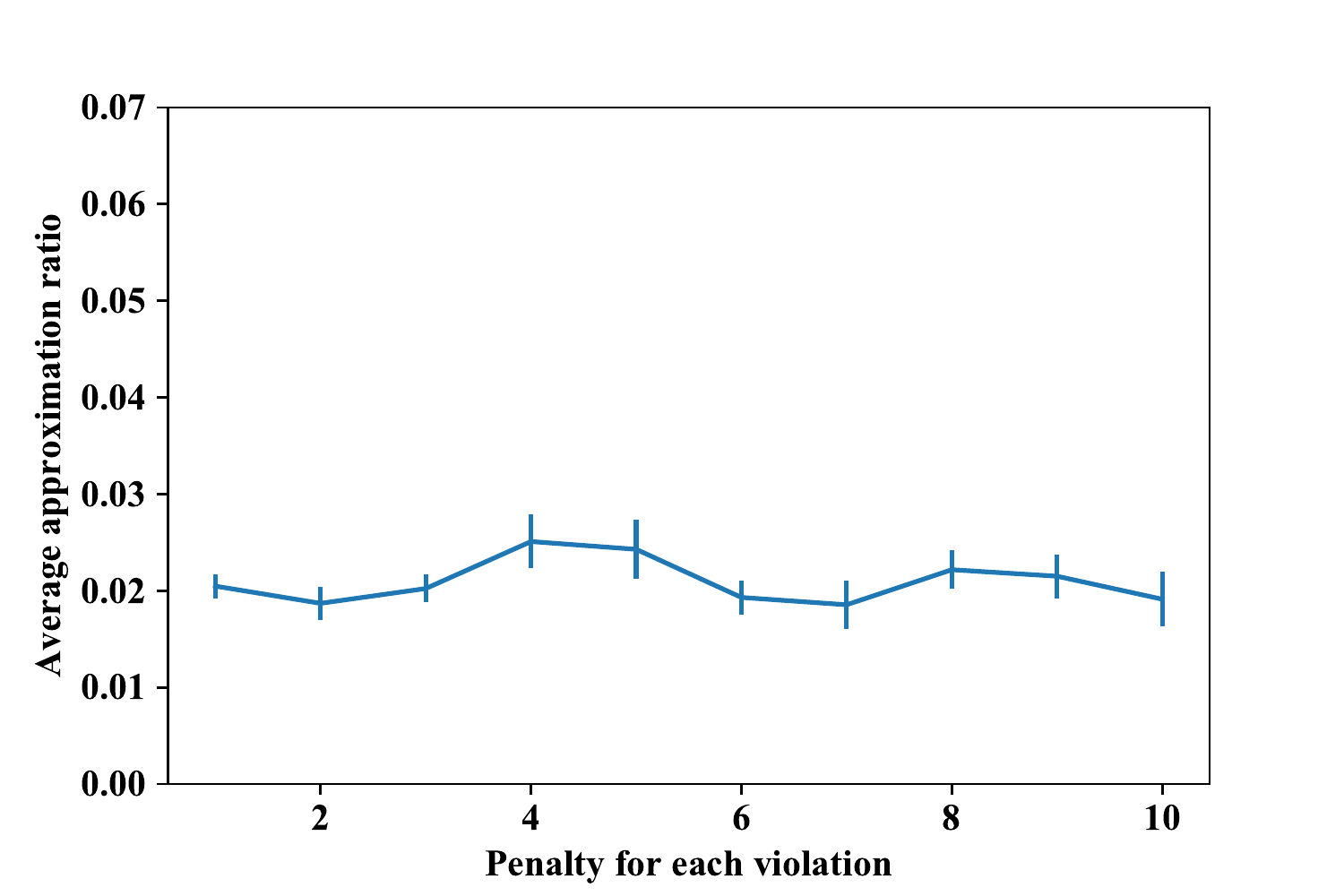}
    \caption{{\bf Behavior of X-mixer QAOA with different penalties, $\Delta$} for SSSP problem at $P=1$. This result shows that the average behavior of X-mixer QAOA is fairly insensitive to the precise choice of the penalty coefficient $\Delta$. }
    \label{fig:Penalty}
\end{figure}

To compare the performance of QAOA on network flow problems using different X-, QED-, and RQED-mixers, we adopt a metric called the approximation ratio (AR)~\cite{wang2020x}, defined as:
\begin{align}
    \text{AR}(\v{\gamma},\v{\beta}) = \frac{\<\psi_p(\v{\gamma},\v{\beta})|\Pi \({C_{\max}-H_C}\)\Pi|\psi_p(\v{\gamma},\v{\beta})\>} {C_{\max}-C_{\min}}
\end{align}
where  $C_{\max}$ and $C_{\min}$ respectively represent the maximum and minimum costs from the set of feasible solutions, and $\Pi$ is the projector into the feasible subspace, which ensures that only states without constraint violations and isolated loops are counted.
The approximation ratio indicates fractional of improvement compared to the worst case, normalized by the possible range of cost values, despite whether an EDP instance on a certain problem exists. In practice, we perform multiple independent runs to obtain average performance, namely, the average approximation ratio (AAR), as the indicator of QAOA performance.
Similarly, the variational optimization of QAOA parameters is done with respect to the projected cost function:
\begin{align}\label{eq:costf_proj}
	\tilde{\e}_C(\v{\gamma},\v{\beta}) := \<\psi_p(\v{\gamma},\v{\beta})|\Pi H_C\Pi|\psi_p(\v{\gamma},\v{\beta})\>.
\end{align}
%
Whereas, by construction, the QED- and RQED-mixers automatically avoid flow-conservation violating constraints, flow-constraint violations can only be softly penalized by introducing an extra term to the cost function for the X-mixer:
\begin{equation}
\label{eq:pen}
	H_{C,\text{penalty}} = \Delta \sum_{u\in\V,i}\left(\sum_{(u,v)\in\E}E_{(u,v)}^{(i)}-d_i(\delta_{u,s_i}-\delta_{u,t_i})\right)^2.
\end{equation}
In principle, $\Delta$ introduces an extra hyperparameter that must be optimized. Generally, $\Delta$ should increase with problem size to avoid the tendency to lower cost by violating constraints. For the problem-sizes we simulate, the results are not very sensitive to the precise choice in $\Delta$ (Figure.~\ref{fig:Penalty}), and we choose $\Delta=1$ throughout for simplicity.

\subsection{Mixer comparison on SSSP problems}
\begin{figure}[h]
    \centering
    \includegraphics[width=0.5\textwidth]{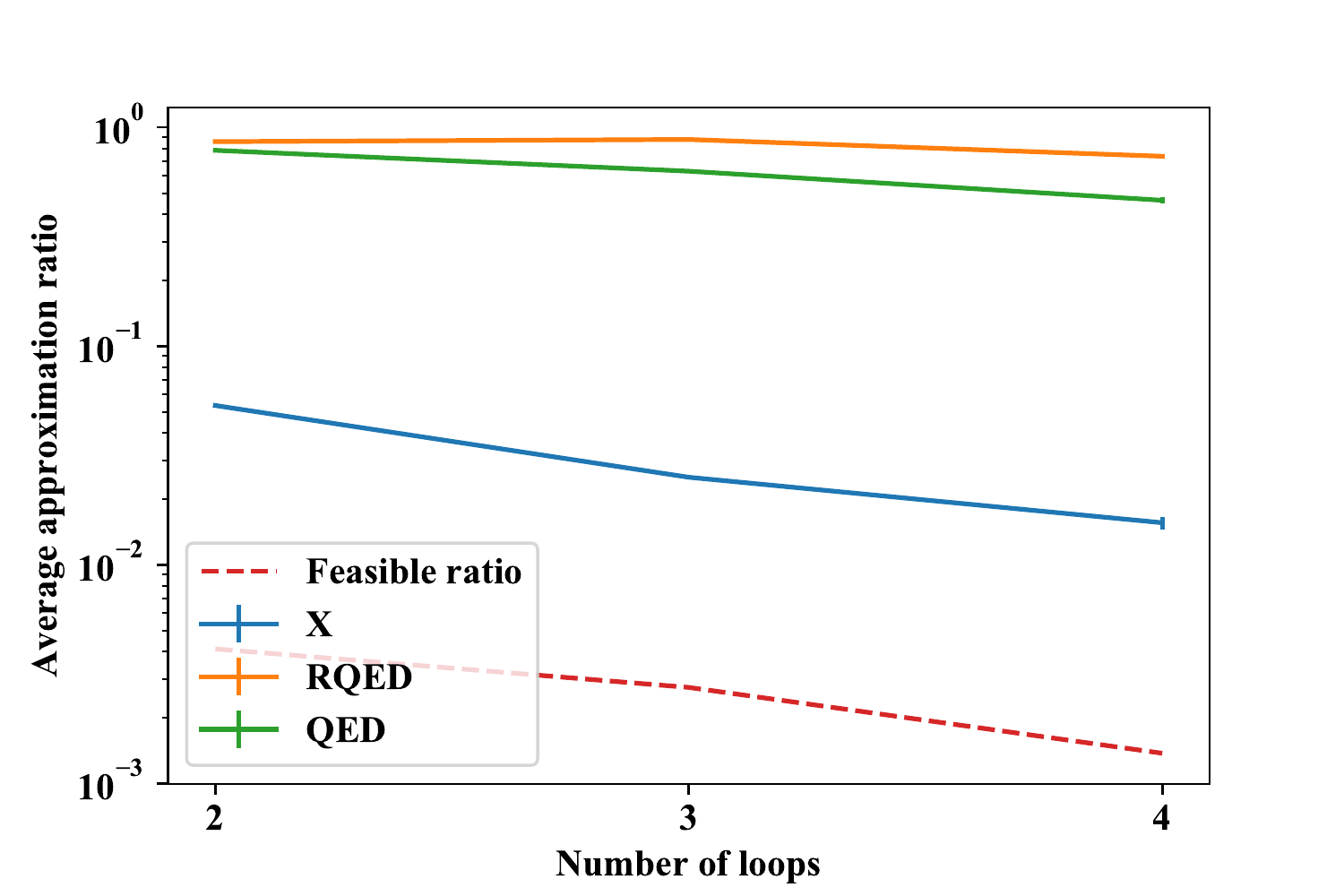}
    \includegraphics[width=0.5\textwidth]{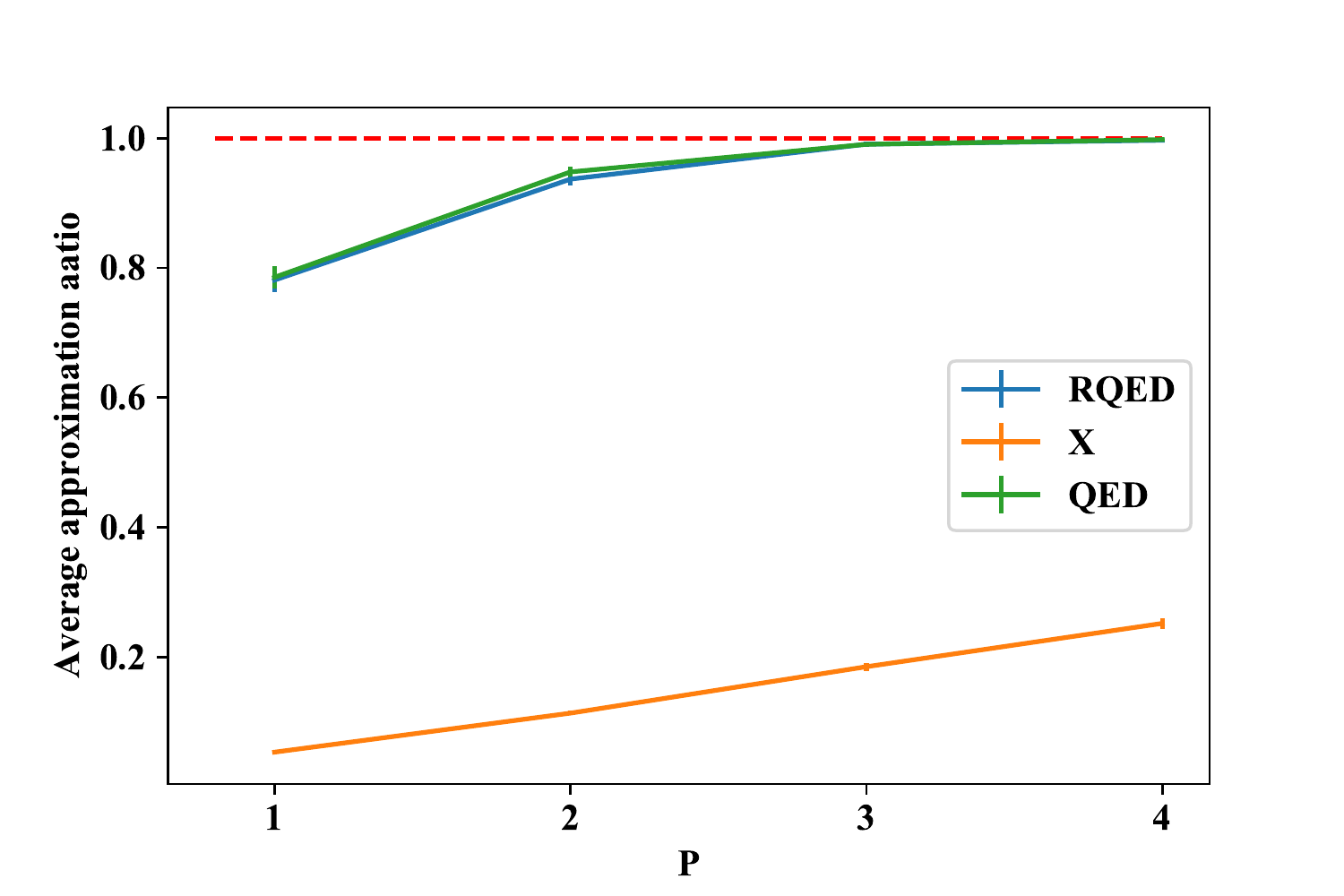}
    \caption{{\bf Comparing different mixers Top:} Solving SSSP on different sized triangle graphs with different mixers; 120 runs performed for each mixer:  weight on each edge is randomly drew from $[0, 1]$.
    {\bf Bottom:} A multiple-step comparison: We compare the behavior of the 3 mixers in solving SSSP problem on the 2-triangle graph as plotted in Figure~\ref{fig:triangle}. Each point represents an average of 200 random instances. Notice that, for this particular graph, it is impossible for the unrestricted QED-mixer to create an isolated loop, making its performance almost identical to the RQED-mixer.}
    \label{fig:mixer-comp}
\end{figure}

We begin with a comparison of the performance between all three mixers: the X-, QED- and RQED-mixer, for approximately solving SSSP problems on different sized graphs. As expected, X-mixer exhibits substantially worse performance than the flow-constraint preserving QED mixers. For a single QAOA round, $p=1$, the degradation in X-mixer's performance with increasing graph sizes tracks the decreasing trend of the ratio between the number of feasible solutions and the size of whole Hilbert space (as shown in Figure.~\ref{fig:mixer-comp}).

The unrestricted QED mixer initially matches the RQED-mixer on the smallest problem instances, for which the graphs are too small to permit isolated loop creation. As the graph size grows, the unrestricted QED mixer's AAR drops below that of the RQED-mixer. For the largest graphs, the QED-mixers' AAR approaches the value achieved by picking feasible paths at random, showing that isolated loop creation can substantially degrade the unrestricted QED-mixer performance at $p=1$.

This shows that, although we start with a feasible solution, isolated loops can be created when using the QED-mixer in its original version. A multi-step QAOA simulation shows that, for the 2-triangle graph, the QED-mixers are able to solve the problem exactly at around $p=3$, which is not surprising due to the small size of the problem.

\subsection{EDP on Undirected Graphs}
Even though a direct comparison between the X-mixer and the QED-mixers for EDP problems is expensive, we test out the performance of the QED-mixers alone on larger graphs by restricting the simulation to the feasible subspace to reduce the computational power required.
In order to be able to compare performances at different graph sizes, we only consider EDP problems with $k=2$ source-sink pairs.
\begin{figure}[h]
    \centering.
    \includegraphics[width=0.5\textwidth]{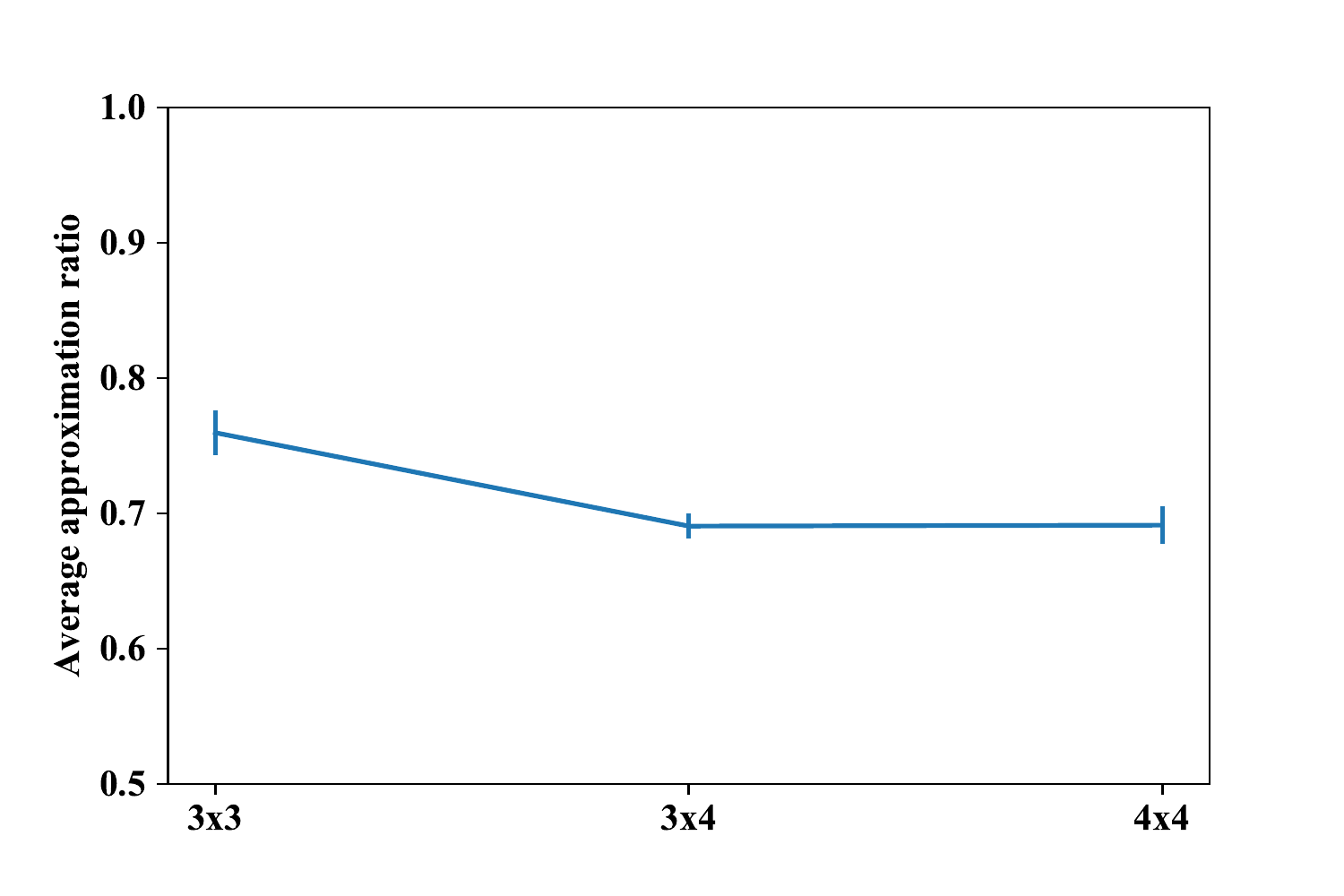}
    \caption{{\bf RQED-Mixer Behavior at $p=1$} The simulation is done for $3\times3$, $3\times4$, $4\times4$ grids for a 2-pair EDP problem. 200 random problem instances are performed at each graph, by choosing the location of each source, sink, state preparation seed path at random.}
    \label{fig:QED}
\end{figure}
As shown in Figure.~\ref{fig:QED}, even though the solution space size for $4\times4$ grid is typically 100 or more times (depending on the location of sinks and sources) than that of the $3\times3$-grid, the performance is only weakly affected -- even after only a single QAOA round, $p=1$, the AAR remains higher than 0.7.
\begin{figure}[h]
    \centering
    \includegraphics[width=0.5\textwidth]{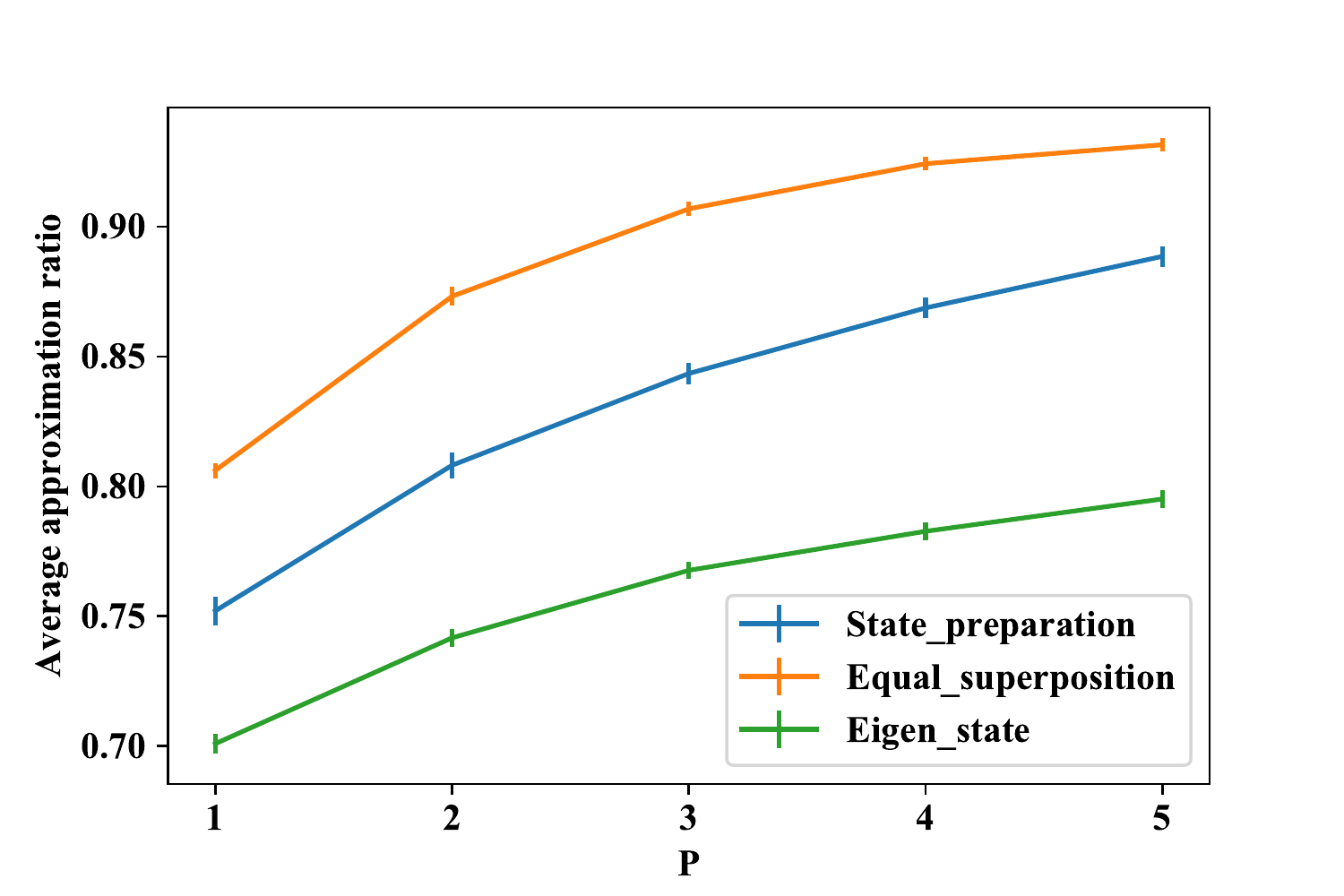}
    \caption{{\bf RQED QAOA behavior in solving actual EDP problems with different initial states} We compare the effect of different choices of initial states on the RQED-mixer's performance, averaging over 200 random problem instances. $\text{``Eigenstate"}$ stands for the ground state.}
    \label{fig:edp}
\end{figure}
As a complementary to the results in IPR test, Figure.~\ref{fig:edp}, shows how different initial state IPRs result in different outcomes in solving EDP on $3\times3$ grids. We observe that, the ground state preparation is not a necessity for our mixer, but the equal superposition state of all feasible solutions does serve as a best starting point of the three, followed closely by the initial state prepared by evolving a random configuration with the mixing Hamiltonian, which is with the IPR test.

These results suggest that having an unbiased ergodic superposition of solutions is more advantageous than starting close to the mixer ground-state (for ordinary QAOA with the X-mixer, these coincide). 


\section{Discussion}
In this work, we designed and simulated a QED-inspired QAOA algorithm to the flow network problems. In particular, we tested its performance with the EDP and SSSP problems. The biggest difference between routing problems and other typical QAOA benchmark problems (like MaxCut) is that the feasible solutions only consist of an exponentially small fraction of the whole solution space. The standard QAOA approach produces infeasible solutions with high probability. To resolve this issue, we proposed the RQED-mixer, which automatically ensures the satisfaction of flow constraints throughout the algorithm. By observing the analogy between Gauss' Law and those constraints, we theoretically and numerically demonstrated that the QED-mixer is a natural choice for the routing problem. 
Although implementing the RQED-mixer requires additional circuit complexity, the generating Hamiltonian is still local and the number of terms is still linear in problem size, and optimization purely within the feasible space makes the QAOA with RQED-mixer more likely to find nearly optimal solutions in comparison to the standard QAOA approach. Part of the simulation results showed that for SSSP problem, the average approximation ratio of RQED-mixer is significantly higher than the X-mixer. For the harder problem, EDP, our results also showed that QAOA with RQED-mixer can achieve high approximation ratio on different size instances, although our numerical simulations were necessarily limited to rather modest problem sizes. 

Our experiments with different initial state strategies suggest an intriguing departure from the ``shortcut-to-adiabaticity" mechanism typically used to motivate QAOA. Namely, QAOA is often motivated as a short-depth approximation to the adiabatic mapping from mixer to the cost of ground-state. However, we have seen that, at least on modest graph sizes available for classical simulation, starting with a more ergodic (less biased) superposition of initial states produces better results than starting in a low-energy state of the mixer, suggesting that a different mechanism than approximate adiabaticity is at play.

Whether the improved performance and superiority of the non-adiabatic operations extend to a larger problem size is an important question for future studies. However, the scope of classical simulation is limited due to the typical explosion of Hilbert space size with problem size. Analytic insights would be extremely valuable, although have often proved challenging beyond small-$p$. One possible approach is to investigate the locality of QAOA with the RQED-mixer. For standard QAOA with X-mixer, the locality was studied \cite{farhi2020quantum} to prove the performance of QAOA on the independent set problem, another famous NP-complete problem on graph. Last but not least, it would be desirable to implement the algorithm on near-term quantum computers, as these devices begin to eclipse classical simulation \cite{48651}.

\section*{Acknowledgements} We thank David Hayes for insightful conversations. The authors acknowledge the Texas Advanced Computing Center (TACC) at The University of Texas at Austin for providing high-performance computing resources that have contributed to the simulation results reported within this paper. URL: http://www.tacc.utexas.edu. This work was supported by NSF DMR-1653007 (AP) and NSF Grant CCF1648712 (RZ).

\appendix
\section{Proof of the QED-mixer's universality for planar graphs \label{app:completeness}}
 In this section we prove that QED-mixer is able to evolve between two arbitrary paths in $O(n)$ loop operations on a planar graph.
\emph{Given a undirected graph $G(V,E)$ with $k$ pairs $(s_i,t_i)$. The goal is to find $k$ paths connecting $s_i$ and $t_i$ for all $i\in [k]$ such that the maximum of congestion in each edge is minimized.}

\emph{Proof:} We first assume that $P_1$ and $P_2$ do not have any common vertex. Then, $s\xrightarrow{P_1} t\xrightarrow{-P_2} s$ forms a closed simple region, where $-P_2$ means the inverse direction of path $P_2$. By Jordan's theorem \cite{hal07}, we can take the ``interior" of this region, which is a subgraph $G'$ of $G$. It's easy to see that every cycle in $G'$ is also a cycle in $G$. Hence, we can apply a loop operation for every cycle in $G'$, and doing so in some specific directions will transform $P_1$ to $P_2$. Wlog., suppose $s\xrightarrow{P_1} t\xrightarrow{-P_2} s$ is in clockwise direction. Then, for every cycle, we apply a counter-clock loop operator, which is equivalent to transfer 1 unit of flow counter-clockwise through the cycle. Let $e=(u,v)$ be an edge in $G'$. If $e$ is contained in $P_1$ and initially there is 1 unit flow from $u$ to $v$. After the loop operation, another 1 unit flow from $v$ to $u$ is introduced so that the total flow on $e$ is 0. Similarly, if $e$ is contained in $P_2$ and is in the same direction as $P_2$, then the flow on $e$ is 1. For all the interior edges, the flow on them is 0 because each edge is contained in two cycles and the loop operation on each cycle will introduce 1 unit flow through $e$ in opposite directions, which will be cancelled by each other. Therefore, after these loop operations, the flow from $s$ to $t$ through $P_1$ will be transformed to $P_2$.

In general, let $v_1=s,\dots, v_l=t$ be $l$ common vertices between $P_1$ and $P_2$, sorted by their appearance orders in the path. Then, we can see that for all $i\in [t-1]$, $v_i\rightarrow v_{i+1}\rightarrow v_{i}$ forms a closed simple region and we take the subgraph $G'_i$. For each $G'_i$, we can apply a series of loop operations to transform $v_{i}\xrightarrow{P_1} v_{i+1}$ to $v_i \xrightarrow{P_2} v_{i+1}$. Therefore, after processing $t-1$ subgraphs, $P_1$ will be transformed to $P_2$. 

Lastly, we show that the number of loop operations we applied is $O(n)$. For each cycle, we only apply the corresponding loop operation once. Hence, the number of loop operations is upper bounded by the number of cycles in $G$. Since $G$ is planar, Euler characteristic for planar graph gives $n-m+f=2$, where $m$ is the number of edges in $G$ and $f$ is the number of cycles. Thus, we have $f=m-n+2$. We also know that, for planar graph, $m\leq 3n-6$. Hence, $f\leq 2n-4=O(n)$. Therefore, we can transform $P_1$ to $P_2$ by $O(n)$ loop operations.
\section{Dual ``height-model" formulation}

Here we consider a canonical (for a detailed description, see, ~\cite{Fisher2018Chapter1D}, for example) dual picture description of the algorithm on plane graphs that might be useful for implementation sometimes. In graph theory, the dual of any plane graph $G$ is obtained by taking each of its faces as a vertex, and drawing an edge between any two neighboring faces. In this dual representation, an initial configuration is chosen, and the states are defined by the relative ``loop distance" to the initial configuration. The amount of flow on each path is then equal to the initial flow plus the difference between states of adjacent faces with the direction perpendicular counterclockwise to the gradient direction. Namely, a ``1" state on some elementary loop adds a counterclockwise flow loop to the initial configuration, and vice versa.
\label{app:dual}
 \begin{figure}[h]
    \centering.
    \includegraphics[scale=0.3]{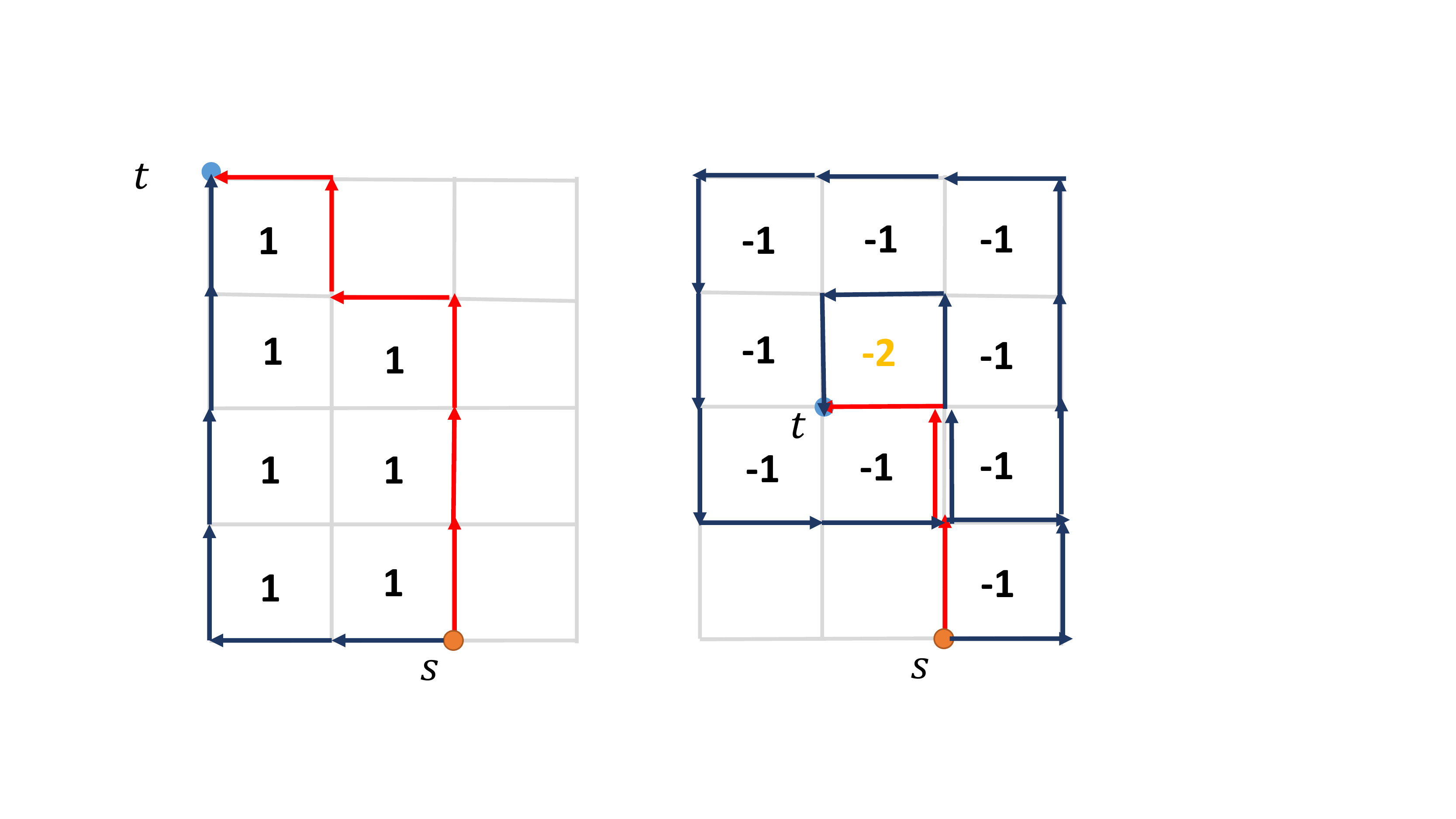}
    \caption{{\bf Examples of dual picture description} We consider a single-pair flow instance: red represents initial/reference configuration, and blue stands for the final configuration. Numbers on each faces stands for the states encoded in dual picture language; every unlabeled face has state 0. In some cases (left) the range of dual state could be simply ${-1, 0, 1}$ whereas more complicated paths (right) needs greater range, which could be proportional to the radius of the graph, namely $O(\sqrt{n})$.}
    \label{fig:dual}
\end{figure}
Naively, one would think that the total Hilbert size for EDP problem becomes $3^{kf}$, where $f$ stands for the number of faces; this makes the Hilbert space a polynomial order less than the original picture, considering $e>f$ on planar graphs. In addition, we could prepare equal superposition states in the dual picture. Nevertheless, there are cases which require more levels on each face, as shown in Figure.~\ref{fig:dual}. For larger graphs, the dual encoding thus becomes even more expensive. On the other hand, the encoding still cannot get rid of isolated loops, although a direct interpretation of RQED-mixer is possible. In conclusion, the dual description can be a useful alternative when considering the ordinary QED-mixer on small graphs, but adds significant qubit resource overheads for larger graphs.


\bibliographystyle{plainnat}

\bibliography{main.bib}

\end{document}